\documentclass[runningheads]{llncs}
\usepackage{graphicx}

\usepackage[colorlinks]{hyperref}
\usepackage{tikz}
\usepackage{xcolor}
\begin{document}

%more space for tables
%\renewcommand{\arraystretch}{1.4}
\setlength{\tabcolsep}{0.3em}

%
%\title{Semantic Segmentation of 4D cardiac CT sequences.}
\title{4D CNN for semantic segmentation of cardiac volumetric sequences}
\titlerunning{4D CNN for semantic segmentation of cardiac volumetric sequences}
% \author{Anonymous}
\author{Andriy Myronenko\inst{1} \and
        Dong Yang\inst{1} \and
        Varun Buch\inst{2} \and
        Daguang Xu\inst{1} \and
        Alvin Ihsani\inst{1} \and
        Sean Doyle\inst{2} \and
        Mark Michalski\inst{2} \and
        Neil Tenenholtz\inst{2} \and
        Holger Roth\inst{1}
        }
% %
\authorrunning{Myronenko et al.}
% \institute{***\\ ***\\ ***}
\institute{NVIDIA \email{\{amyronenko,dongy,daguangx,aihsani,hroth\}@nvidia.com}
\and MGH and BWH Center for Clinical Data Science  \email{\{varun.buch,sdoyle\}@mgh.harvard.edu}
\email{\{mmichalski1,ntenenholtz\}@partners.org}
}

\maketitle              % typeset the header of the contribution
\vspace{-1.5\baselineskip}

\begin{abstract}
We propose a 4D convolutional neural network (CNN) for the segmentation of retrospective ECG-gated cardiac CT, a series of single-channel volumetric data over time. While only a small subset of volumes in the temporal sequence is annotated, we define a sparse loss function on available labels to allow the network to leverage unlabeled images during training and generate a fully segmented sequence. 
We investigate the accuracy of the proposed 4D network to predict temporally consistent segmentations and compare with traditional 3D segmentation approaches. We demonstrate the feasibility of the 4D CNN and establish its performance on cardiac 4D CCTA\footnote[1]{video: \url{https://drive.google.com/uc?id=1n-GJX5nviVs8R7tque2zy2uHFcN_Ogn1}}. 
\end{abstract}

\section{Introduction}
\label{sec:intro}

%1) describe the medical problem, is there a cardiac disease? why do we need segmentation?
%2) describe CTA

Cardiovascular disease is responsible for 18 million deaths annually, making it one of the leading causes of mortality globally \cite{who2017cardio}. Coronary computed tomography angiography (CCTA) uses contrast-enhanced CT to evaluate cardiac muscle morphology, function, and vascular patency. Two measurements derived from CCTA with significant diagnostic and prognostic importance are the Left Ventricular Ejection Fraction (LVEF) and Left Ventricular Wall Thickness. Both measurements require the segmentation of the left ventricular muscle, with the former requiring temporal segmentation over the cardiac cycle. The American College of Radiology (ACR) has highlighted the importance of these measurements by listing them among the most important initial `use cases' of artificial intelligence as applied to radiology \cite{touch_ai_directory_2019}. A segmentation model of the left ventricular muscle and cavity over the cardiac cycle, especially the end-systole and end-diastole time points, would allow for automated determination of both measurements from 4D CCTA studies. The clinical utility of such a model is highly relevant as it reduces study reading time and improves the consistency of measurements, thereby potentially preventing missed pathology in cases where the measurements may not have otherwise been performed. 

Modern 4D CCTA images are acquired over the entire cardiac cycle, including end-systole and end-diastole. A typical 4D scan includes 20 3D volumes reflecting the cardiac anatomy at equally-spaced time points within a 240 ms time interval. This allows for enough temporal resolution to study the heart's function. In order to limit the amount of effort required to annotate these images, we restrict the annotation to only certain frames, an example of which is shown in Fig.~\ref{fig:seg}. 

While convolutional neural networks (CNNs) have demonstrated state-of-the-art performance across a variety of segmentation tasks~\cite{Ronneberger15}, the adoption of 4D CNNs for 4D medical imaging (3D + time -- e.g., CT or ultrasound) has been limited due to the high computational complexity and lack of manually segmented data. The cost of annotating volumetric imaging is significant, making 4D labeling prohibitively expensive. Nevertheless, the temporal dimension offers valuable information that is otherwise lost when treating each volume independently.

In this work, we propose a 4D CNN for the segmentation of the left ventricle (LV) and left ventricular myocardium (LVM) from 4D CCTA images, enabling the computation of the aforementioned cardiac measurements. To reduce annotation costs, our 4D dataset is sparsely labeled across the temporal dimension -- only a fraction of volumes in the sequence are labeled. This enables us to leverage a 4D CNN with a sparse loss function, allowing our algorithm to take advantage of unlabeled images which would otherwise be discarded in a 3D model. The network jointly segments the sequence of volumes, implicitly learning temporal correlations and imposing a soft temporal smoothness constraint. We describe the 4D convolution layer generalization in Section~\ref{sec:4dconv} and introduce a sparse Dice loss function as well as a temporal consistency regularization in Section~\ref{sec:loss}. We demonstrate the feasibility of a 4D CNN and compare its performance to a traditional 3D CNN in Section~\ref{sec:results}.

\section{Related work}
\label{sec:relatedwork}

Deep learning has achieved state-of-the-art segmentation performance in 2D natural images~\cite{deeplabv3plus2018} and 2D \cite{Ronneberger15} \& 3D medical images~\cite{Milletari16,Myronenko18}. 
To leverage the temporal dependency and account for segmentation continuity, recurrent neural networks (RNNs) have been adopted for videos~\cite{valipour2017recurrent} and 2D+T cardiac MRI datasets~\cite{zhang2018segmentation}.  %Convolutional models can also represent temporal relationships and offer competitive performance for language translation~\cite{gehring2017convolutional}. 
3D CNNs have also been applied spatio-temporally and proven effective in segmentation of videos~\cite{tran2015learning,tran2016deep} and 2D+T cardiac MRIs~\cite{yang2018multi}.

For sequences of volumetric imaging, such as 3D+T CT or ultrasound, 4D CNNs are a natural extension. Wang et al.~\cite{wang2016dataset} proposed a CNN for 4D light-field material recognition incorporating separable 4D convolutions to reduce computational complexity. 
Clark et al.~\cite{clark2019convolutional} adopted a 4D CNN for the de-noising of low-dose CT, where three independent 3D convolutions (with fixed cyclic time delay) were used to simulate 4D convolutions. 

To date, 4D CNNs for semantic segmentation have not been explored in similar depth to 2D and 3D CNNs, in part due to their high computational requirements and lack of available annotations. In this work, we demonstrate the feasibility and advantageousness of a true 4D CNN.

\section{Methods}
\label{sec:methods}

\begin{figure}[t] 
 \centering
 \includegraphics[clip=true, trim=0pt 0pt 0pt 0pt, width=0.6\textwidth]{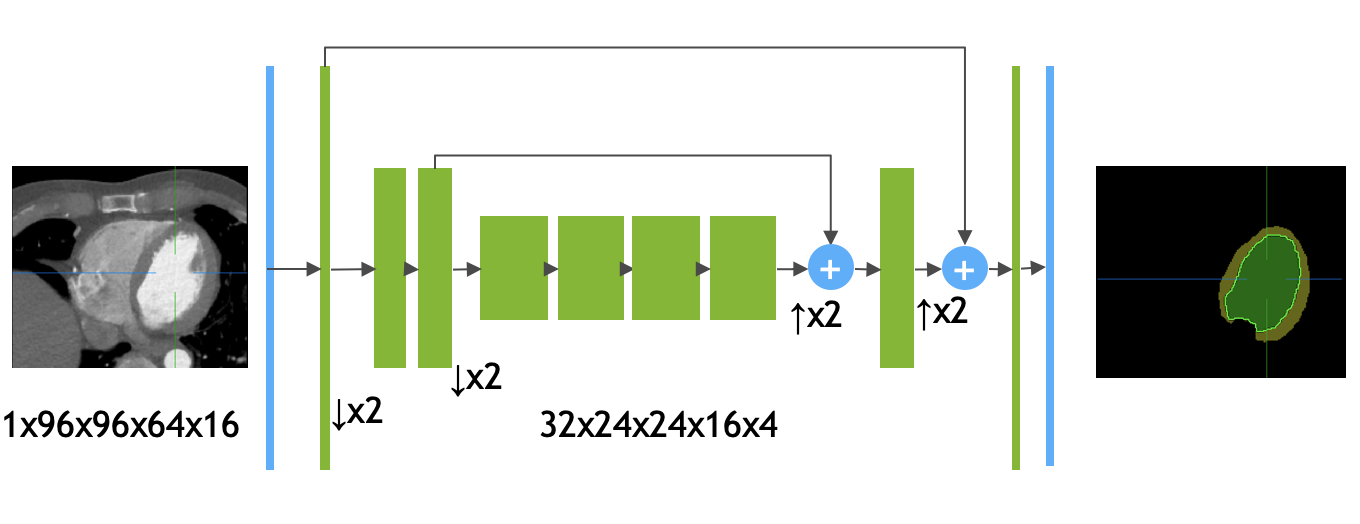}
 \caption{4D network architecture:  input is a single channel (grayscale) 4D CT crop, followed by initial 3x3x3x3 4D convolution with 8 filters. Each green building block is a ResNet-like block with GroupNorm normalization. The output has three channels followed by a softmax: background, left ventricle, and myocardium. For a detailed description of the building blocks see Table~\ref{tab:network}.}
 \label{fig:network}
 \vspace{-5mm}
\end{figure}

Our 4D segmentation network architecture follows an encoder-decoder semantic segmentation strategy, typical for 2D and 3D images. Throughout the network, we use 4D convolutions with a kernel size of 3x3x3x3, where the last dimension corresponds to time.  The network architecture follows the one proposed in~\cite{Myronenko18}, where only the main decoder branch is used and modified to fit 4D images within GPU memory limits. 
The input size of the network is 1x1x96x96x64x16 (corresponding to a batch size of 1, input channel 1, and a spatial crop of 96x96x64 with 16 frames). We randomly crop this 4D array from the input data during training. No other form of augmentation is employed in this study. 

Each building block of the network consists of two convolutions with group normalization~\cite{Wu18} and ReLU, followed by identity skip-connections similar to ResNet~\cite{He16} blocks. % Group Normalization (GN)~\cite{Wu18} is used for normalization.   
A sequence of the building blocks is applied sequentially at different spatial levels. In the encoder part of the network, we downscale the spatial dimension after each level and double the feature dimension. We use strided convolutions (stride of 2) for downsizing, and all convolutions are 3x3x3x3. We use one block at level 0 (initial size), two blocks at level 1, and four blocks at level 2. At the smallest scale, the input image crop is downsized by a factor of 4 (to 24x24x16x4), which provides a balance between network depth and GPU memory limits.
For the encoder branch, we leverage a similar structure with a single block per each spatial level. To upsample, we use 4D nearest-neighbor interpolation after 1x1x1x1 convolution. Finally, we use additive skip-connections between the corresponding levels. The details of network structure are shown in Table~\ref{tab:network} and in Fig.~\ref{fig:network}.

\begin{table}
	\small

	\centering
	\caption{Network structure, where GN stands for group normalization (with group size of 8), Conv - 3x3x3x3 convolution, AddId - addition of identity/skip connection. Repeat column shows the number of repetitions of the block. The output, after softmax, has 3 channels (background and 2 foreground classes)}
	\label{tab:network}
	\begin{tabular}{|l|cc|c|}
		 \hline
		Name & Ops & Repeat &Output size    \\ \hline
		Input & &  &1x96x96x64x16    \\
		InitConv & Conv 3x3x3x3 &  & 8x96x96x64x16    \\
		EncoderBlock0 & GN,ReLU,Conv,GN,ReLU,Conv, AddId &  &8x96x96x64x16    \\
		EncoderDown1 & Conv 3x3x3x3 stride 2 & & 16x48x48x32x8    \\
		EncoderBlock1 & GN,ReLU,Conv,GN,ReLU,Conv, AddId& x2 &16x48x48x32x8    \\
		EncoderDown2 & Conv 3x3x3x3 stride 2& & 32x24x24x16x4    \\
		EncoderBlock2 & GN,ReLU,Conv,GN,ReLU,Conv, AddId& x4 &32x24x24x16x4    \\
	    DecoderUp1 & Conv1, UpNearest, +EncoderBlock1 &   &16x48x48x32x8    \\
		DecoderBlock1 & GN,ReLU,Conv,GN,ReLU,Conv, AddId &  & 16x48x48x32x8    \\
		DecoderUp0 & Conv1, UpNearest, +EncoderBlock0 &   &8x96x96x64x16    \\
		DecoderBlock0 & GN,ReLU,Conv,GN,ReLU,Conv, AddId &  & 8x96x96x64x16   \\
		DecoderEnd & Conv 1x1x1x1, Softmax &   &3x96x96x64x16    \\
		
		\hline
	\end{tabular}
\end{table}

\subsection{4D convolutions}
\label{sec:4dconv}
While 4D convolutional layers are not available in common deep-learning frameworks (such as TensorFlow\footnote{\url{https://www.tensorflow.org}} of PyTorch\footnote{\url{https://pytorch.org}}), they can be represented as a sum over a sequence of 3D convolutions along the fourth (temporal) dimension. For efficiency, we rearranged the loop to avoid repeated 3D convolutions by implementing 4D convolution as a custom TensorFlow layer. This strategy allows for a true (non-separable) 4D convolution.
A common approach to maintain the same image dimension is to zero-pad prior to a convolution. We were concerned that such an approach may introduce boundary effect for the very first and last frames (when padding with zeros). We have experimented with several padding strategies for the 4th dimension only, including zero padding, mirror reflection, and replication but did not observe any noticeable performance differences, thus we decided to use conventional zero padding.

 \subsection{Loss}
 \label{sec:loss}
  %[ToDo] how many labeled images in each sequence?
Our training dataset is sparsely labeled along the temporal dimension since labeling medical images in 4D (and even in 3D) is complex and time-consuming. Therefore, we have defined a sparse loss function that is applied only to the labeled time-frames and includes a regularization term to ensure temporal consistency between frames.

The proposed loss function is therefore composed of two terms, 
 \newcommand{\LL}{\ensuremath{\mathbf{L}}}
\begin{equation}
 \label{eq:loss}
 \LL = \sum_{i \in \mathrm{labeled}} D(p_{\mathrm{true}}^i, p_{\mathrm{pred}}^i ) +  \sum_{i=0}^{K-2} ||p_{\mathrm{pred}}^{i+1} - p_{\mathrm{pred}}^i ||^2  
 \end{equation}
where $D$ is a soft dice loss~\cite{Milletari16} applied only to labeled time points (3D images) $p_{\mathrm{true}}$ to match the corresponding outputs $p_{\mathrm{pred}}$:
\begin{equation}
  \label{eq:dice}
  D(p_{\mathrm{true}},  p_{\mathrm{pred}})  = 1 - \frac{2*\sum p_{\mathrm{true}} * p_{\mathrm{pred}} }{\sum p_{\mathrm{true}}^2 + \sum p_{\mathrm{pred}}^2 + \varepsilon}   
\end{equation}
$K$ is the number of frames (K=16 in our case, since we use the 96x96x64x16 crop size). The second term in (\ref{eq:loss}) is a first-order derivative over time to enforce similarity between frames. Re-weighting the contributions between the loss terms did not show consistent difference, so we kept the equal contributions.

 \subsection{Optimization}
 Similar to~\cite{Myronenko18}, we apply the Adam optimizer with an initial learning rate of $ \alpha_{0} = \mathtt{1}\mathrm{e}{\mathtt{-3}} $ and progressively decrease it according to the following schedule $ \alpha = \alpha_{0} \left(1-\eta/N_{\eta}\right)^{0.9}$, where $\eta$ is an epoch counter, and $N_{\eta}$ is the total number of training epochs.

 We use a batch size of 1 and sample input sequences randomly (ensuring that each training sequence is drawn once per epoch). From each 4D sequence, we apply a random crop of size 96x96x64x16 centered on a foreground (with a probability of 0.6), otherwise centered on a background voxel. Thus, at each iteration, a different number of ground truth labels is available, depending on the location of the crop window (16) of the time dimension.
 
 \subsection{Dataset}
Our dataset consists of 61 4D CCTA sequences, each of 512x512x(40-108)x20 size (512x512 axial size, with 40-108 slices of variable thickness and 20 time points). The spatial image resolution is (0.24-0.46)x(0.24-0.46)x2mm. All images were acquired at Massachusetts General Hospital, Boston, USA, using a 128-slice dual-source multi-detector CT with retrospective ECG gating and tube current modulation. Sequences were reconstructed from multiple R-R\footnote{R corresponds to the peak of the QRS complex in the ECG wave.} intervals, measured via electrocardiogram.

All images were resampled to an isotropic spatial resolution of 1x1x1mm, retaining the temporal resolution. After re-sampling, the 4D image sizes vary between 112x122x80x20 and 238x238x158x20 voxels. We apply a random data split, with 49 4D images used for training and 12 4D images for validation. 

The number of annotated frames in each sequence varies widely, ranging from only 2 out 20 (i.e. end-systole and end-diastole) to 9 (every second time point).  Overall, 247 time-points have been annotated throughout the dataset, which represents approximately 20\% of all frames.
%ranging from 2 (end-systole and end-diastole) to 9 (every second time point) in some cases. 
We include studies with differing numbers of annotated frames in both training and validation splits to maximize temporal coverage during both training and validation.

As a second form of validation, we compare our model's segmentation results with clinical findings. One such clinical finding is the ejection fraction measure which typically is being judged as reduced when less than 55\% \cite{curtis2003association}.

\section{Results}
\label{sec:results}

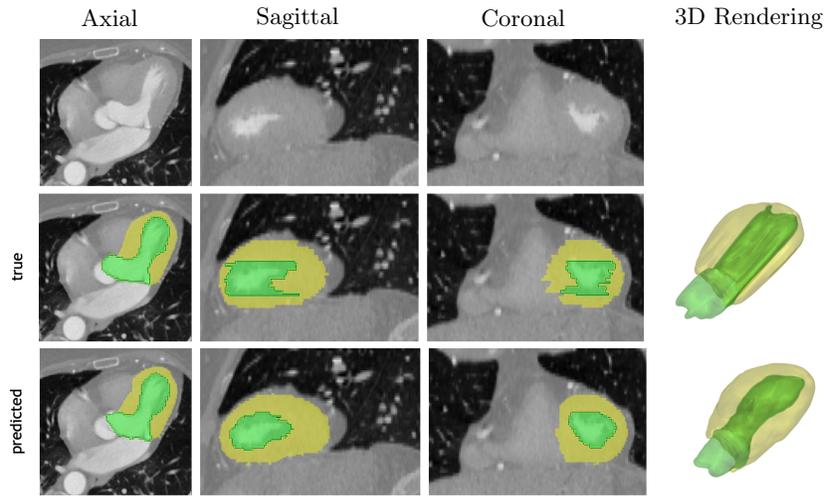
\begin{figure}[t] 
 \centering
 \begin{tikzpicture}
\node at (0,0) { \includegraphics[clip=true, trim=0pt 0pt 0pt 0pt, width=0.9\textwidth]{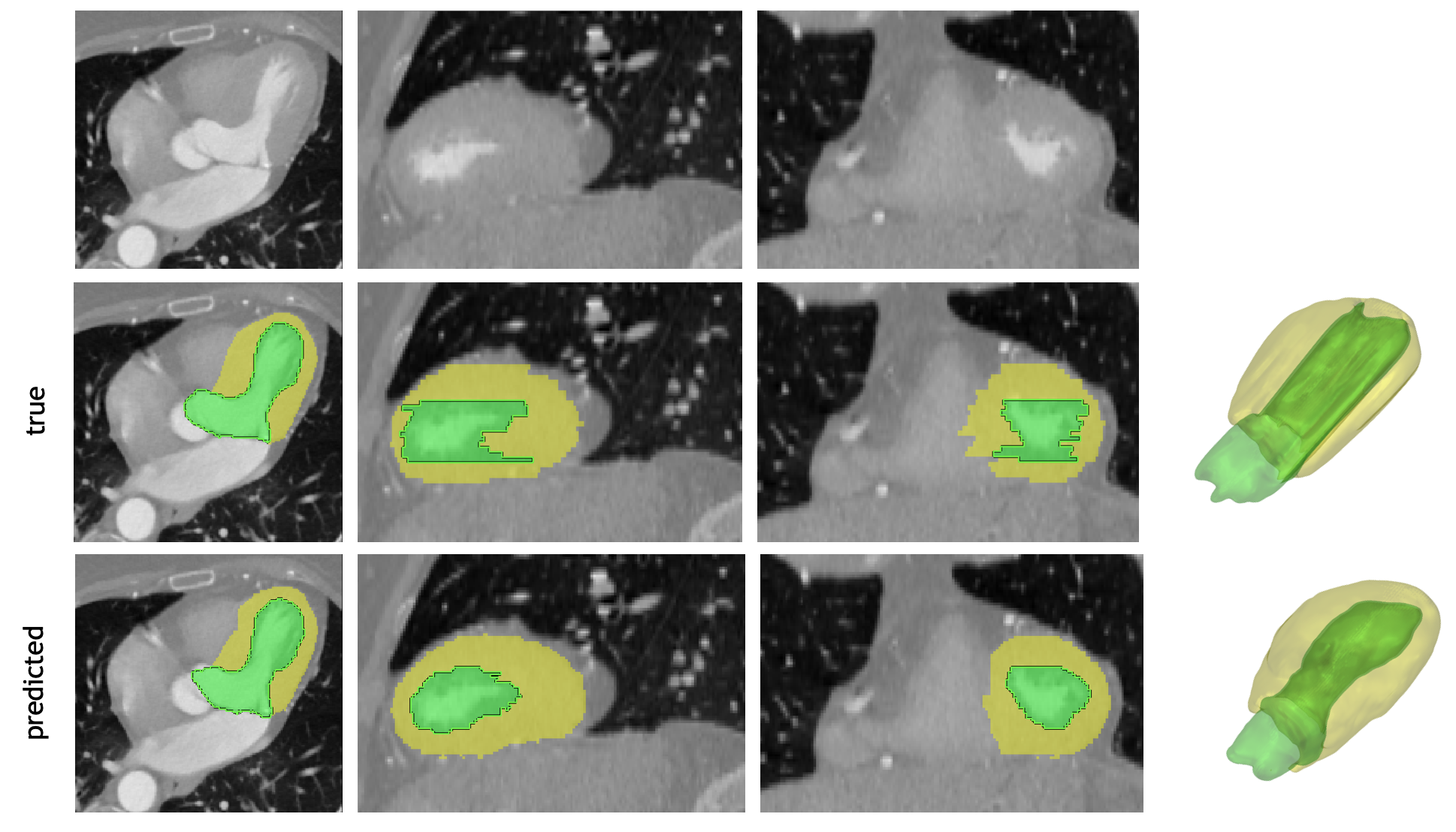} };
\node at (-4, 3.35) {Axial};
\node at (-1.5, 3.35) {Sagittal};
\node at (1.5, 3.35) {Coronal};
\node at (4.5, 3.35) {3D Rendering};
\end{tikzpicture}
 \caption{A typical segmentation example of our 4D network in axial, sagittal and coronal views of a single 3D frame. Notice that the predicted results look better and much smoother than manual annotations in sagittal and coronal cross sections. Manual labeling was done by a trained clinician slice-by-slice, which results in noisy out-of-plane ground-truth labels. The 4D segmentation network is able to average out these errors when learning such noisy data examples.}
 \label{fig:seg}
 \vspace{-5mm}
\end{figure}

%\subsection{Experiments}

We implemented our 4D network in Tensorflow and trained it on an NVIDIA Tesla P100 SXM2 GPU with 16GB memory based on the \textit{NVIDIA Clara Train SDK}\footnote{\url{https://devblogs.nvidia.com/annotate-adapt-model-medical-imaging-clara-train-sdk}}. 
%During training we used a random crop of size 96x96x64x16, batch size 1. 
Data is normalized to [-1,1] using a fixed scaling from input CT range [-1024,1024]. We train for 500 epochs and use the model at the end of training for evaluations. 

For comparison, we also implemented a 3D network largely following the same architecture as in Fig. ~\ref{fig:network}, except that all convolutions are 3D and include a greater number of layers with one additional down-sampling level (the end of the encoder being of size 12x12x8) as GPU memory requirements permit deeper architecture in the 3D case. For the 3D network, we use a crop of size 96x96x64 and train it only on labeled 3D frames. The 3D network learns to predict segmentation without any temporal constraint considerations. We acknowledge that such a 3D network is trained on less number of images (only the annotate frames), and weakly-supervised 3D segmentation might be a candidate for better comparison.

\noindent\emph{Segmentation performance:} We evaluate both networks on the validation set, using only the labeled frames, in terms of average Dice score. In addition, we assess the temporal continuity of the produced results. A temporal smoothness metric, we compute the L2 norm of the first-order time derivative of segmentation labels, as well as the average surface distance between the consecutive frames. 
Intuitively, accurate segmentation results must respect the temporal continuity of the heart motion, and are expected to be smoother in the time domain.

\begin{table}
\vspace{-1mm}

	\centering
	%\small
	\caption{Performance evaluation of the 4D semantic segmentation network. LVM - left ventricular myocardium and LV - left ventricle. We also measure temporal smoothness in the result using the L2 norm of temporal derivative of the predictions and average surface distance between the consecutive frames. The proposed 4D network produces temporally smoother results with comparable dice scores.}
	\label{tab:results}
	\begin{tabular}{|l|c|c|c|c|}
	
	\hline
 & \multicolumn{2}{c|}{Dice} & \multicolumn{2}{c|}{Smoothness}                                        \\ \hline

		Arch  & LVM & LV  & L2 & Surf    \\ \hline
        3D network & 0.85 & 0.91 & 1.28& 0.74 \\
		4D network & 0.85 & 0.90 & 1.05 & 0.59    \\

		\hline
	\end{tabular}
\vspace{-4mm}
\end{table}

The evaluation results are shown in Table~\ref{tab:results}.  
In terms of the dice score alone, the proposed 4D network demonstrated only comparable results, with one of the structures (LV cavity) 1\% better dice of the 3D network. One reason for this might be that 4D network is not as deep as its 3D counterpart and the dice score is estimated frame by frame; frame-by-frame Dice score may not be the most representative accuracy measure of temporal sequence segmentations as it does not account for consistency across frames. 

Visually, the 4D CNN segmentation results have superior temporal consistency, where the label changes more ``fluidly'' between time-frames. Our smoothness metric confirms this observation, with the proposed 4D network achieving lower smoothness loss than its 3D counterpart (see Table~\ref{tab:results}).  
We also observe that in many cases, 4D CNN results look better than the ground truth (See Fig.~\ref{fig:seg}). The manual annotations are done slice-by-slice, which results in jittery out-of-plane annotation profiles; this especially visible in sagittal and coronal views. The proposed 4D segmentation network is able to average out these errors while learning from the overall dataset and produce coherent results both spatially and temporally. In future work, manual relabeling of some cases in all 2D planes consistently (in spatial and time dimensions) could result in a clearer advantage of our 4D approach.  

\noindent\emph{Ejection fraction:} We computed the ejection fraction for 12 cases (10 with normal and 2 with reduced ejection fraction) based on the ratio of minimum and maximum LV cavity volume throughout the cardiac cycle as predicted by our models. For both, 3D and 4D models, we achieve a 100\% sensitivity and specificity in detecting reduced ejection fraction when compared to the findings reported in the clinical reports (provided by radiologists).

\section{Conclusion}
\label{sec:conclusion}

We proposed a 4D convolutional neural network for semantic segmentation of the left ventricle (LV) and left ventricular myocardium (LVM) from 4D CCTA studies. The network is fully convolutional and jointly segments a temporal sequence of volumetric images from CCTA.

We utilize a sparse Dice loss function and a temporal consistency regularization to handle the problem of sparse temporal annotation. We have demonstrated the feasibility and advantageousness of a true 4D CNN compared to 3D CNNs, where the first shows improvement in segmentation temporal consistency. The model's result showed promise in being useful for automatically quantifying clinically measures, such as ejection fraction. 

%TODO

\bibliographystyle{splncs04}
\bibliography{paper}

\begin{thebibliography}{10}
\providecommand{\url}[1]{\texttt{#1}}
\providecommand{\urlprefix}{URL }
\providecommand{\doi}[1]{https://doi.org/#1}

\bibitem{touch_ai_directory_2019}
{American College of Radiology}: Touch-ai directory (2019),
  \url{https://www.acrdsi.org/DSI-Services/TOUCH-AI}

\bibitem{deeplabv3plus2018}
Chen, L.C., Zhu, Y., Papandreou, G., Schroff, F., Adam, H.: Encoder-decoder
  with atrous separable convolution for semantic image segm. arXiv:1802.02611
  (2018)

\bibitem{clark2019convolutional}
Clark, D., Badea, C.: Convolutional regularization methods for 4d, x-ray ct
  reconstruction. In: Medical Imaging: PMI. vol. 10948 (2019)

\bibitem{curtis2003association}
Curtis, J.P., Sokol, S.I., Wang, Y., Rathore, S.S., Ko, D.T., Jadbabaie, F.,
  Portnay, E.L., Marshalko, S.J., Radford, M.J., Krumholz, H.M.: The
  association of left ventricular ejection fraction, mortality, and cause of
  death in stable outpatients with heart failure. ACC  \textbf{42}(4),
  736--742 (2003)

\bibitem{He16}
He, K., Zhang, X., Ren, S., Sun, J.: Identity mappings in deep residual
  networks. In: European Conference on Computer Vision (ECCV) (2016)

\bibitem{Milletari16}
Milletari, F., Navab, N., Ahmadi, S.A.: V-net: Fully convolutional neural
  networks for volumetric medical image segmentation. In: Fourth International
  Conference on 3D Vision (3DV) (2016)

\bibitem{Myronenko18}
Myronenko, A.: {3D} {MRI} brain tumor segmentation using autoencoder
  regularization. In: {BrainLes}, {MICCAI}. pp. 311--320. LNCS, Springer
  (2018), \url{https://arxiv.org/abs/1810.11654}

\bibitem{Ronneberger15}
Ronneberger, O., P.Fischer, Brox, T.: U-net: Convolutional networks for
  biomedical image segmentation. In: Medical Image Computing and
  Computer-Assisted Intervention (MICCAI). LNCS, vol.~9351, pp. 234--241.
  Springer (2015)

\bibitem{tran2015learning}
Tran, D., Bourdev, L., Fergus, R., Torresani, L., Paluri, M.: Learning
  spatiotemporal features with {3D} convolutional networks. In: Proceedings of
  the IEEE international conference on computer vision. pp. 4489--4497 (2015)

\bibitem{tran2016deep}
Tran, D., Bourdev, L., Fergus, R., Torresani, L., Paluri, M.: Deep end2end
  voxel2voxel prediction. In: Proceedings of the IEEE conference on computer
  vision and pattern recognition workshops. pp. 17--24 (2016)

\bibitem{valipour2017recurrent}
Valipour, S., Siam, M., Jagersand, M., Ray, N.: Recurrent fully convolutional
  networks for video segmentation. In: 2017 IEEE Winter Conference on
  Applications of Computer Vision (WACV). pp. 29--36. IEEE (2017)

\bibitem{wang2016dataset}
Wang, T.C., Zhu, J.Y., Hiroaki, E., Chandraker, M., Efros, A., Ramamoorthi, R.:
  A {4D} light-field dataset and {CNN} architectures for material recognition.
  In: Proceedings of European Conference on Computer Vision (ECCV) (2016)

\bibitem{who2017cardio}
{World Health Organization}: Cardiovascular diseases ({CVDs}) (May 2017),
  \url{https://www.who.int/en/news-room/fact-sheets/detail/cardiovascular-diseases-(cvds)}

\bibitem{Wu18}
Wu, Y., He, K.: Group normalization. In: European Conference on Computer Vision
  (ECCV) (2018)

\bibitem{yang2018multi}
Yang, D., Huang, Q., Axel, L., Metaxas, D.: Multi-component deformable models
  coupled with 2d-3d u-net for automated probabilistic segmentation of cardiac
  walls and blood. In: ISBI. pp. 479--483 (2018)

\bibitem{zhang2018segmentation}
Zhang, D., Icke, I., Dogdas, B., Parimal, S., Sampath, S., Forbes, J., Bagchi,
  A., Chin, C.L., Chen, A.: Segmentation of left ventricle myocardium in
  porcine cardiac cine mr images using a hybrid of fully convolutional neural
  networks and convolutional lstm. In: Medical Imaging 2018: Image Processing.
  vol. 10574, p. 105740A. International Society for Optics and Photonics (2018)

\end{thebibliography}

\end{document}